\documentclass[twocolumn,aps,pra]{revtex4}
\usepackage{amsmath}
\usepackage[dvips,xdvi]{graphicx}
\usepackage{amssymb}
\usepackage{epsfig}
\usepackage{dcolumn}
\newcommand{\be}{
\begin{equation}
}
\newcommand{\ee}{
\end{equation}
}
\newcommand{\ba}{
\begin{array}}
\newcommand{\ea}{
\end{array}}
\newcommand{\bea}{
\begin{eqnarray}
}
\newcommand{\eea}{
\end{eqnarray}
}
\newcommand{\beas}{
\begin{eqnarray*}
}
\newcommand{\eeas}{
\end{eqnarray*}
}

\begin{document}
\title{Ansatz from Non-Linear Optics Applied to Trapped Bose-Einstein Condensates}
\author{Murat Ke\c{c}eli, F. \"{O}. Ilday, M. \"{O}. Oktel}
%\email{keceli@fen.bilkent.edu.tr}
\affiliation{ Department of Physics, Bilkent University, 06800 Ankara, Turkey }
\date{\today}
\begin{abstract}

A simple analytical ansatz, which has been used to describe the
intensity profile of the similariton laser \cite{il04,ji06} is
used as a variational wave function to solve the Gross-Pitaevskii
equation for a wide range of interaction parameters. The
variational form interpolates between the non-interacting density
profile and the strongly interacting Thomas-Fermi profile
smoothly. The simple form of the ansatz is modified for both
cylindrically symmetric and completely anisotropic harmonic traps.
The resulting ground state density profile and energy are in very
good agreement with both the analytical solutions in the limiting
cases of interaction and the numerical solutions in the
intermediate regime.

\end{abstract}
\maketitle

It is common for dynamical systems with weak coupling to show
Gaussian behavior with respect to key parameters, whereas strongly
coupled and highly nonlinear systems tend to exhibit power law
dependencies. A striking recent example of this is the ultrashort
pulse formation in the similariton laser~\cite{il04}. Another
nonlinear system, which exhibits a parabolic density profile for
strong interactions is a trapped Bose-Einstein Condensate (BEC).
In Bose-Einstein condensation, the density of the condensate is
analogous to the intensity of light in nonlinear optics and the
nonlinear governing equation for this system, which is called the
Gross-Pitaevskii equation (GPE) is very similar to the equation
for propagation of laser light in a nonlinear optical medium.
Based on this similarity, it is natural to expect similar
solutions for these vastly different systems. Soliton-like and
self-similar solutions of nonlinear Schrodinger Equation (NLSE)
are important both in BEC and NLO.

Soliton-like solutions arise when the nonlinearity is compensated
by the dispersion, and they are the only exact analytical
solutions of these NLSEs, whereas self-similar solutions are
asymptotic solutions that show up when the effects of initial
conditions die out but the system is still far from the final
state \cite{ba96}. Although soliton type solutions have been
extensively studied in both the NLO and BEC communities,
self-similar solutions are not as comprehensively investigated. In
optics, these type of solutions are used more extensively from
Raman scattering to pulse propagation in fibers, and it is shown
that linearly chirped parabolic pulses are exact asymptotic
solutions of NLSE with gain~\cite{SS}. Recently, we have
demonstrated experimentally and numerically that self-similar
propagation of ultrashort parabolic pulses (similaritons) are
stable in a laser resonator~\cite{il04}. More recently, we have
developed a semi-analytic theory of the similariton
laser~\cite{ji06,ji07}. Instrumental in this step was the
introduction of an ansatz to describe the intensity profile of
this pulse, which can be `tuned' to any condition, ranging from
weakly nonlinear (Gaussian pulses) to strongly nonlinear
(parabolic pulses). Here, motivated by the mathematical similarity
between the two systems, we apply the same ansatz to describe the
density profile of a BEC in a quadratic trap. We show that this
ansatz describes the system with excellent accuracy, in the whole
range from the non-interacting limit to the strongly interacting
limit.

Gross-Pitaevskii theory \cite{gp61} gives a very successful description of the ground state and excitations of the BECs
in dilute atomic gases. The success of this theory implies that the condensate can be described accurately with a single
wave function and the interactions between the particles are through  $s$-wave scatterings. The interaction of the
particles are then represented by the interaction strength $g=\frac{4\pi\hbar^2a}{m}$, where $a$ is the $s$-wave
scattering length, and $m$ is the mass of the trapped particles. The theory reduces to a single equation that describes
the condensate wave function, known as GPE; a type of nonlinear Schr\"{o}dinger equation which arises in many areas of
physics like NLO and hydrodynamic theory of fluids.

\begin{table}[htbp]\label{table1}
\begin{center}\caption{The values of the wave function at the center, the root mean square sizes $r_{rms}$ and chemical
potentials are tabulated in units of $\sqrt{N/a_w^3}$, $a_w$ and $\hbar\omega$ respectively. For comparison numerical
results of Ref. \cite{bt03} are given in parentheses.}
\begin{tabular}{|l|l|l|l|}
\hline
\multicolumn{1}{|c|}{$\beta$} & \multicolumn{1}{c|}{$\psi(0)$} & \multicolumn{1}{c|}{$r_{rms}$} &
\multicolumn{1}{c|}{$\mu$} \\
\hline 0 & \multicolumn{1}{c|}{0.4238 (0.4238)   } & 1.2248 (1.2248) & 1.5000 (1.5000) \\
0.2496 & 0.3969 (0.3843) & 1.2794 (1.2785) & 1.6805 (1.6774) \\
0.9986 & 0.3475 (0.3180) & 1.3981 (1.3921) & 2.0885 (2.0650) \\
2.4964 & 0.2515 (0.2581) & 1.5355 (1.5356) & 2.5803 (2.5861) \\
9.9857 & 0.1739 (0.1738) & 1.8822 (1.8821) & 4.0089 (4.0141) \\
49.926 & 0.1097 (0.1066) & 2.5071 (2.5057) & 7.2576 (7.2484) \\
249.64 & 0.0665 (0.0655) & 3.4152 (3.4145) & 13.559 (13.553) \\
2496.4 & 0.0330 (0.0328) & 5.3855 (5.3852) & 33.812 (33.810) \\
\hline
\end{tabular}
\end{center}
\end{table}

\begin{figure}\centering
\includegraphics[scale=0.5]{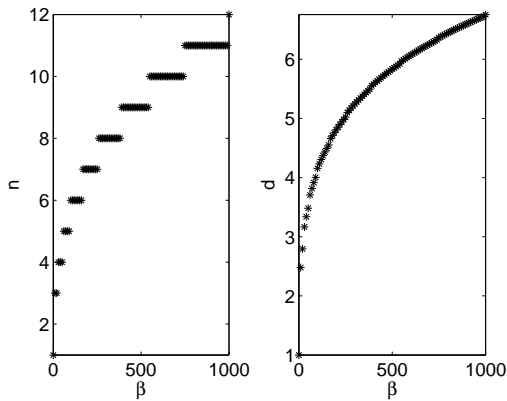}
\caption{Change of variational parameters with interaction
parameter $\beta$ is given. Left plot shows the number of terms in
the summation and the right one shows the change of width of the
similariton ansatz in units of oscillator length
$a_\omega$.}\label{dn} % birim
\end{figure}

 GPE can be obtained by minimizing the ground state energy functional of the condensate
\begin{equation}
\label{energy}
    E(\psi)=\int
d\textbf{r}\left(\frac{\hbar^2}{2m}|\nabla\psi(\textbf{r})|^2+V(\textbf{r})|\psi(\textbf{r})|^2+\frac{g}{2}|\psi(%
\textbf{r})|^4\right).
\end{equation}
with respect to the wave function. The terms in the energy functional correspond to kinetic, trapping and interaction
energies, respectively. The trapping potential can generally be approximated with a harmonic potential for many of the
experiments. Time-independent GPE follows as,
\be
\label{GPE}
-\frac{\hbar^2}{2m}\nabla^2\psi(\textbf{r})+V(\textbf{r})\psi(\textbf{r})+g|\psi(\textbf{r})|^2\psi(\textbf{r})=\mu%
\psi(%
\textbf{r}),
\ee where $\mu$ is chemical potential introduced as the Lagrange multiplier for the normalization constraint $\int
d\textbf{r}n(\textbf{r})=N$, where $n(\textbf{r})=|\psi(\textbf{r})|^2$ is the density of the condensate.

\begin{figure}\centering
\includegraphics[scale=0.5]{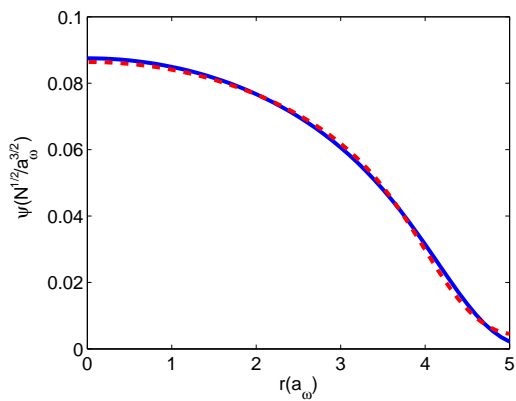}
\caption{(Color online) Wave function calculated with the steepest
descent method \cite{ds96} is shown with bold line (blue) whereas
the similariton ansatz solution is given with dashed line (red)
for $\beta=100$. }\label{simvscomp}
\end{figure}

Nonlinearity of GPE is due to interaction between particles and
its effect becomes more pronounced as the number of particles in
the condensate increases which is the case for current experiments
where more than $10^7$ particles form
the BEC. Since very few exact solutions of NLSE, such as solitons,
are known , many numerical algorithms
\cite{eb95,rh95,ds96,sf99,bt03} and variational methods
\cite{bp96,im96,pg97,fe97,be98,ss99} are developed to find ground
state solutions. Although variational methods give only an upper
bound to the ground state energy, they require less calculation
and can give accurate results if a suitable trial function is
chosen. Another advantage of the variational principle is that it
gives the functional form of the wave function which can be used
to obtain other properties of the condensate.

\begin{figure}\centering
\includegraphics[scale=0.5]{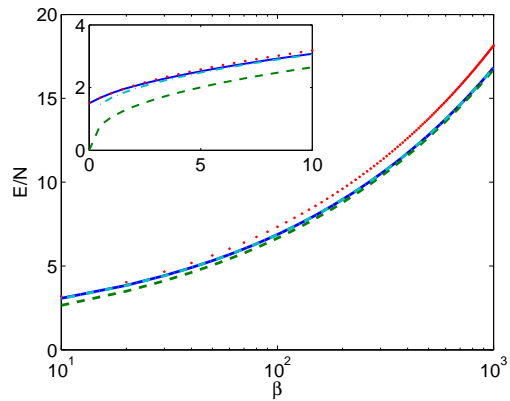}
\caption{(Color online) Ground state energy with respect to
interaction parameter $\beta$ obtained with the variational
function (black solid line). The resulting energy of a Gaussian
variational function is given with dotted (red) line and energy
obtained with Thomas-Fermi solution is given with dashed line
(green). Improved Thomas-Fermi solution \cite{lp96} is given with
dotted line (blue). Energy per particle is given in units of
$\hbar\omega$. The inset is given for small $\beta$ values. }
\label{energyiso}
\end{figure}

Therefore, many trial functions are proposed to obtain a lower
bound for the ground state energy. Trial functions are generally
chosen by adding parameters to a known approximate analytical
solution. The trivial approximate solutions are obtained by
looking at the limiting cases of the GPE where the nonlinearity is
negligibly small or very high. Assuming an ideal Bose gas equation
(\ref{GPE}) reduces to the Schr\"{o}dinger equation where the
chemical potential corresponds to the energy eigenvalue, by
neglecting the nonlinear term. With a harmonic trap the problem
turns into a harmonic oscillator problem where the solution is a
Gaussian. For the opposite case where nonlinearity is dominant,
the kinetic energy term can be neglected in GPE, and then the
equation can easily be solved for the density,$
n(\textbf{r})=\frac{\mu-V(\textbf{r})}{2g}$,when the right hand
side of the equation is positive. This approximation is known as
Thomas-Fermi approximation (TFA) and it shows that as the
interaction increases, the density profile changes from a Gaussian
to a parabola for a harmonic trap. TFA can be improved by adding
the kinetic energy term obtained with the resulting wave function
with a suitable cut off \cite{lp96} as, \be \label{ekin}
\frac{E_{kin}}{N}=(15\beta)^{-2/5}(\frac{1}{2}\ln{(480\beta)}-\frac{5}{4}).
\ee

With this insight many trial functions are proposed, for isotropic traps and anisotropic traps with cylindrical symmetry
to describe the intermediate regime where neither TFA and nor the ideal gas approximation is valid. Here, we make use of
a simple analytic function which has already been used in NLO where a similar behavior -Gaussian to parabolic- for
intensity profile of a similariton laser.

The so-called `similariton' pulse in optics has a nearly parabolic intensity profile to reduce the effect of Kerr
nonlinearity. However, if the nonlinearity of the system is reduced, the pulse assumes the well-known Gaussian shape of
the dispersion-managed solitons. Therefore, the ansatz proposed in
\cite{ji06} to describe these pulses has an adjustable profile between a Gaussian and an inverted parabola
\begin{equation}
\label{sim}
    S_n(x)=\exp{(-\sum_{k=1}^n{\frac{x^{2k}}{k}})}.
\end{equation}
This function becomes a Gaussian when $n=1$ \be S_{n=1}(x)=
\exp{(-x^2)}, \ee and turns into an inverted parabola when $n$
goes to $\infty$ since the summation in the exponent converges to
$\ln(1-x^2)$ for $|x|<1$, \be S_{n \rightarrow \infty} =
\exp{(-\sum_{k=1}^\infty{\frac{x^{2k}}{k}})}= \exp{[\ln(1-x^2)]}=
1-x^2. \ee Moreover the expansion converges so quickly that adding
about ten terms is enough to get a parabolic profile with smooth
ends. Besides, this function is easily integrable which makes it a
good candidate for variational calculations.

\begin{table*}[htbp]\label{table2}
\begin{center}\caption{Results of our calculation for a cylindrically harmonic trap with $\lambda=\sqrt{8}$. Energy and
length units are $N\hbar\omega$ and $a_\omega$. The results of the numerical calculation in Ref. \cite{ds96} are given
in parentheses for comparison except for the last row. For $\beta=2165$ Thomas-Fermi result for the chemical potential
is given in parentheses.}
\begin{tabular}{|l|l|l|l|l|l|l|}
\hline
\multicolumn{1}{|c|}{$\beta$} & \multicolumn{1}{c|}{$x_{rms}$} & \multicolumn{1}{c|}{$z_{rms}$} &
\multicolumn{1}{c|}{$E_{kin}$} & \multicolumn{1}{c|}{$E_{tr}$} & \multicolumn{1}{c|}{$E_{int}$} &
\multicolumn{1}{c|}{$\mu$} \\
\hline 0.0000 & 0.7071 (0.707)& 0.4204 (0.42) & 1.2071 (1.207)& 1.2071 (1.207)& 0.0000 (0.000)& 2.4142 (2.414) \\
0.4330 & 0.7901 (0.79) & 0.4374 (0.44) & 1.0539 (1.06) & 1.3894 (1.39) & 0.2237 (0.21) & 2.8907 (2.88) \\
0.8660 & 0.8500 (0.85) & 0.4472 (0.45) & 0.9976 (0.98) & 1.5225 (1.52) & 0.3500 (0.36) & 3.2200 (3.21) \\
2.1650 & 0.9657 (0.96) & 0.4707 (0.47) & 0.8528 (0.86) & 1.8188 (1.81) & 0.6440 (0.63) & 3.9596 (3.94) \\
4.3300 & 1.0892 (1.08) & 0.4966 (0.50) & 0.7337 (0.76) & 2.1730 (2.15) & 0.9595 (0.96) & 4.8258 (4.77) \\
8.6600 & 1.2319 (1.23) & 0.5332 (0.53) & 0.6709 (0.66) & 2.6549 (2.64) & 1.3227 (1.32) & 5.9712 (5.93) \\
21.650 & 1.4798 (1.47) & 0.5930 (0.59) & 0.5314 (0.54) & 3.5963 (3.57) & 2.0432 (2.02) & 8.2142 (8.14) \\
43.300 & 1.7038 (1.69) & 0.6536 (0.65) & 0.4351 (0.45) & 4.6121 (4.57) & 2.7847 (2.74) & 10.616 (10.5) \\
64.950 & 1.8447 (1.84) & 0.6989 (0.70) & 0.4128 (0.41) & 5.3569 (5.31) & 3.2960 (3.26) & 12.361 (12.2) \\
86.600 & 1.9562 (1.94) & 0.7319 (0.73) & 0.3789 (0.38) & 5.9693 (5.91) & 3.7270 (3.68) & 13.802 (13.7) \\
2165 & 3.7367 & 1.3297 & 0.1459 & 21.035 & 13.926 & 49.033 (48.329)\\
\hline
\end{tabular}
\end{center}
\end{table*}

Motivated by these properties, we  use the similariton ansatz  Eq. \ref{sim} as our trial wave function to minimize the
energy functional given in Eq. (\ref{energy}). To simplify the calculations, we non-dimensionalize the Gross-Pitaevskii
functional  by scaling length, energy and wave function with oscillator length
$a_{\omega}=\sqrt{\frac{\hbar}{m\omega}}$, $\hbar\omega$ and $\sqrt{Na_{\omega}}$, respectively. We first analyze the
solution for spherical harmonic trap $V(r)=\frac{1}{2}m\omega^2r^2$ and introduce the parameter $\beta\equiv
Na/a_\omega$ which is a measure of the strength of the interaction. With this rescaling the energy functional becomes
\begin{equation}
\label{scaledenergy}
    \frac{E(\psi)}{N}=2\pi\int_0^\infty \!
d^3r\left(|\nabla\psi({r})|^2+V(r)|\psi({r})|^2+2\pi\beta|\psi({r})|^4\right).
\end{equation}
Ideally $\beta$ can take any value between $-\infty$ to $\infty$ since all the parameters are experimentally tunable.
However, negative scattering length which means attractive interaction, causes collapse of the condensate when the
particle number is high. In this regime, our results agree with \cite{bp96}. In the present work, we concentrate on
repulsive interaction. With proper normalization the trial wave function has the form,
\be\label{simwf}
    \psi(r)=\sqrt{\frac{1}{4\pi d^3I}}\exp(\sum_{k=1}^n{\frac{(r/d)^{2k}}{2k}}),
\ee where $d$ and $n$ are our variational parameters with $I=\int_0^\infty dr r^2 S_n(r)$ which is an integral that can
be calculated analytically for $n=1,2$ and numerically for $n>2$. Here the parameter $d$ is responsible for the width of
the condensate which increases as the interaction increases, and $n$ takes care of flattening of the central density. We
minimize the energy with respect to $d$ for different $n$ values and chose the $n$ that gives the minimum energy. For
$d$, we obtain a fifth order polynomial equation where only one of the roots is physically meaningful.

We compare our results with the analytical approximations as follows. For small $\beta$ values, our trial function
reduces to a Gaussian and gives the exact result for $\beta=0$, and for large $\beta$ our results agree well with the
improved TFA results as shown in Fig. \ref{energyiso}. We also compare the resulting wave function with the numerical
solutions obtained by steepest descent method for different $\beta$ values in Fig. \ref{simvscomp}. We also tabulate our
results in
\ref{table1} and include the results of a recent numerical analysis which minimizes the energy functional directly by
the finite element method. Here it should be noted that tabulated kinetic, trap, and interaction energies satisfy the
virial theorem $2E_{kin}+2E_{pot}-3E_{int}=0$, to our numerical accuracy. It is also remarkable that even for large
$\beta$ adding $10$ terms is enough to find a good approximation for the wave function (see Fig. \ref{dn}) which shows
the simplicity of the calculations.

Using similar trial functions, we can also solve the GPE for a cylindrical trap and a fully anisotropic trap. For the
cylindrically symmetric trap, trial function takes the form,
\be
\label{ansimwf}
\psi(\rho,z)=C\exp(-\sum_{k=1}^{n_\rho}{\frac{(\rho/d_{\rho})^{2k}}{2k}})\exp(-\sum_{k=1}^{n_z}{%
\frac{(z/d_z)^{2k}}{2k}}), \ee where, $C=\sqrt{\frac{N}{2\pi
{d_{\rho}}^2 d_{z} I_{\rho}I_z}}$,$I_{\rho}=\int_0^\infty \rho
d\rho S_n(\rho)$ and $I_z=\int_{-\infty}^\infty dz S_n(z)$. We
have four variational parameters, but calculations are similar to
the isotropic case. We compare our results with the numerical
results of Dalfovo \textit{et al}. \cite{ds96} in Table 2.
Cylindrically symmetric traps are the most common traps in BEC
setups and aspect ratio obtained from $\frac{x_{rms}}{z_{rms}}$ is
very important to identify the BEC phase in these experiments. It
is  shown in \cite{bp96,ds96} that for the noninteracting case
this ratio is equal to $\sqrt{\lambda}$, and goes to $\lambda$ in
the Thomas-Fermi limit. This result is clearly seen from the
values in Table 2 where $\lambda=\sqrt{8}$ and it is also evident
that convergence of TFA is very slow.

There are also experiments with fully anisotropic traps \cite{kd99} and for this case the trial function can also be
modified similarly with six variation parameters. The results are given in Table 3, where agreement with the results in
\cite{sf99} is apparent.

\begin{table}[htbp]\label{table3}\caption{The chemical potentials per particle in units of $\hbar\omega$ are calculated
using an ansatz modified for a completely anisotropic trap with $\lambda=\sqrt{2}$ and $g\gamma=2$. The values of the
interaction parameter $\beta$ are obtained from Ref. \cite{kd99} and the values in parentheses corresponds to the
numerical solution given in \cite{sf99}. The variational parameters $n_{x,y,z}$ and $d_{x,y,z}$ are given.}
\begin{center}
\begin{tabular}{|l|l|l|l|l|}
\hline
\multicolumn{1}{|c|}{$\beta$} & \multicolumn{1}{c|}{$n_x , d_x$} & \multicolumn{1}{c|}{$n_y , d_y$} &
\multicolumn{1}{c|}{$n_z , d_z$} & \multicolumn{1}{c|}{$\mu$} \\
\hline 0 & 1 , 1.000 & 1 , 0.840 & 1 , 0.707 & \multicolumn{1}{c|}{2.207 (2.207)} \\
1.787 & 2 , 1.753 & 2 , 1.368 & 1 , 0.817 & \multicolumn{1}{c|}{3.604 (3.572)} \\
3.575 & 2 ,1.956 & 2 , 1.489 & 2 , 1.160 & \multicolumn{1}{c|}{4.385 (4.345)} \\
7.151 & 3 , 2.433 & 2 ,1.654 & 2 , 1.258 & \multicolumn{1}{c|}{5.492 (5.425)} \\
14.302 & 3 , 2.780 & 3 , 2.035 & 2 , 1.384 & \multicolumn{1}{c|}{7.010 (6.904)} \\
28.605 & 4 , 3.310 & 3 , 2.30 & 3, 1.687 & 9.049 (8.900) \\
57.211 & 5 , 3.880 & 4 , 2.725 & 3 , 1.896 & \multicolumn{1}{c|}{11.78 (11.57)} \\
\hline
\end{tabular}
\end{center}
\end{table}

In summary, an ansatz is introduced to investigate the ground state properties of a BEC at zero temperature for
quadratic traps with arbitrary anisotropy. The ground state energy and wave function are found to be very accurate
anywhere from the noninteracting case to the highly repulsive one, as compared with numerical studies. The form of the
trial function changes from a Gaussian to a parabola smoothly, and it successfully describes the intermediate regime of
moderate interaction. Important quantities like aspect ratio, chemical potential, and root mean square size of the
clouds are calculated and compared to numerical studies \cite{ds96,sf99,bt03}. With a slight modification the suggested
form of the wave function can be applied to vortex states of the condensate. Time-dependent GPE can also be solved using
a similar form to study growth dynamics.

We have shown previously that a nonlinear system in optics, namely a high-energy femtosecond laser oscillator, exists
stably between two extreme limits, corresponding to Gaussian pulse profiles for weak nonlinearity and parabolic profiles
for strong nonlinearity. This indeed appears to be a common behavior observed in many systems, including the trapped BEC
analyzed in this paper. We have a simple analytical function, which has the crucial property of interpolating any state
between these extremes. In this paper, we have shown that the variational approach with the same ansatz yields excellent
results for the BEC system in a quadratic trap. We believe that our approach can be generally applicable to other
nonlinear systems in disparate fields.

\end{document}